\documentclass[usenatbib,referee]{pasa}

\title[The UTMOST: A hybrid digital signal processor transforms the 
  MOST.]{The UTMOST: A hybrid digital signal processor transforms the
  Molonglo Observatory Synthesis Telescope.}

\author[M. Bailes et al.]{\parbox{\textwidth}{M.  Bailes$^{1,2}$,
    A. Jameson$^{1,2}$, C. Flynn$^{1,2}$, T.  Bateman$^{3,4}$,
    E. D. Barr$^{1,2,5}$, S. Bhandari$^{1,2}$, J. D. Bunton$^{4}$, M.
    Caleb$^{6,1,2}$, D. Campbell-Wilson$^{3,2}$, W. Farah$^1$,
    B. Gaensler$^{3,2,9}$, A. J. Green$^{3,2}$, R. W. Hunstead$^{3}$,
    F. Jankowski$^{1,2}$, E. F. Keane$^{7,1,2}$, V. Venkatraman
    Krishnan$^{1,2}$, Tara Murphy$^{2,3}$, M. O'Neill$^1$,
    S. Os{\l}owski$^{1}$, A. Parthasarathy$^{1,2}$, V. Ravi$^{1,2,8}$,
    P. Rosado$^1$, D. Temby$^{3}$} \\ \\ \\
\parbox{\textwidth}
{$^1$Centre for Astrophysics and Supercomputing, Swinburne University of Technology, Mail H30, PO Box 218, VIC 3122, Australia\\
 $^2$ARC Centre of Excellence for All-Sky Astrophysics (CAASTRO),\\
 $^3$Sydney Institute for Astronomy, School of Physics A28, The University of Sydney, NSW 2006, Australia\\
 $^4$Australia Telecope National Facility, PO Box 76, Epping NSW 1710, Australia.\\
 $^5$ Max-Planck-Institut f\"ur Radioastronomie (MPIfR), Auf dem H\"ugel 69, D-53121 Bonn, Germany\\
 $^6$Research School of Astronomy and Astrophysics, Australian National University, Cotter Road Weston Creek, ACT 2611, Australia\\
 $^7$SKA Organization, Jodrell Bank Observatory, Cheshire SK11 9DL, UK\\
 $^8$Cahill Center for Astronomy and Astrophysics, Caltech, Pasadena, CA 91125, USA\\
 $^9$Dunlap Institute for Astronomy and Astrophysics, The University of Toronto, Toronto, ON M5S 3H4, Canada
}}

\usepackage{graphicx}
\usepackage[space]{grffile}
\usepackage{latexsym}
\usepackage{amsfonts,amsmath,amssymb}
\usepackage{url}
\usepackage{upgreek}
\usepackage[utf8]{inputenc}
\usepackage{hyperref}
\hypersetup{colorlinks=false,pdfborder={0 0 0}}
\usepackage{textcomp}
\usepackage{longtable}
\usepackage{multirow,booktabs}

\usepackage[authoryear]{natbib}
\bibpunct{(}{)}{;}{a}{}{,}
\usepackage{aas_macros}
\usepackage{hyperref} 
\hypersetup{colorlinks,citecolor=blue,linkcolor=blue,urlcolor=blue}

\begin{document}

\begin{abstract}
The Molonglo Observatory Synthesis Telescope (MOST) is an 18,000 m$^2$
radio telescope situated some 40 km from the city of Canberra,
Australia. Its operating band (820-850 MHz) is now partly allocated to
mobile phone communications, making radio astronomy particularly
challenging.  In this paper we describe how the deployment of new
digital receivers (RX boxes), Field Programmable Gate Array (FPGA)
based filterbanks and server-class computers equipped with 43 GPUs
(Graphics Processing Units) has helped transform MOST into a versatile
new instrument (the UTMOST) for studying the dynamic radio sky on
millisecond timescales, ideal for work on pulsars and Fast Radio
Bursts (FRBs).  The filterbanks, servers and their high-speed,
low-latency network form part of a hybrid solution to the
observatory's signal processing requirements.  The emphasis on
software and commodity off-the-shelf hardware has enabled rapid
deployment through the re-use of proven 'software backends' for much
of its signal processing needs. The new RX boxes have ten times the
bandwidth of the original MOST and double the sampling of the
cylindrical reflector line feed, which doubles the field of view.  The
servers' record and playback capability has facilitated rapid
debugging of many new observing modes.  Inherent to the solution is
the capacity to run commensal science operations. The UTMOST can
simultaneously excise interference, make maps, coherently dedisperse
pulsars, and perform real-time searches of coherent fan beams for
dispersed single pulses. Although system performance is still
sub-optimal, a pulsar timing and FRB search programme has commenced
and the first UTMOST maps have been made. The telescope operates as a
robotic facility, deciding how to efficiently target pulsars and how
long to stay on source, via feedback from real-time pulsar
folding. The regular timing of over 300 pulsars has resulted in the
discovery of 7 pulsar glitches and 3 FRBs. The UTMOST demonstrates
that if sufficient signal processing can be applied to the voltage
streams it is possible to perform innovative radio science in hostile
radio frequency environments.
\end{abstract}


\maketitle

\label{firstpage}

\section{Introduction} 

The UTMOST\footnote{UTMOST is not an acronym} is a major upgrade of
the Molonglo Observatory Synthesis Telescope (MOST), which consists of
two EW-aligned cylindrical paraboloid reflectors with a collecting
area of 18,000 m$^2$, situated in a valley 40 km east of Australia's
capital city, Canberra. The length of each cylindrical arm is 778 m,
separated by a 15-m gap. It is a synthesis interferometer which was
originally part of the transit instrument known as the Mills Cross
\citep{Robertson_1991}, and is sensitive to right-hand (IEEE)
circularly polarised radiation only. The telescope played a pivotal
role in the early years of radio pulsar astronomy due to its large
collecting area and compact design. Discoveries included the first
association of a pulsar (in Vela) with a supernova remnant
\citep{Large_1968} and 155 new pulsars from the prolific second
Molonglo pulsar survey \citep{Manchester_1978}.

The MOST re-used some of the electronics and analogue systems from the
Molonglo Cross and became a synthesis imaging facility. It processed a
single 3-MHz channel of data streamed from each of 88 18.29-m elements
of the telescope, operating at a central frequency of 843 MHz. Images
were produced by the back projection of fan beams formed in real time
using an analogue, hard-wired beamformer.

As a synthesis array, the MOST produced many important surveys and
catalogues, including the Sydney University Molonglo Sky Survey
\citep[SUMSS;][]{Bock_1999,Mauch_2003}, two Galactic Plane Surveys,
MGPS-1 \citep[]{1999ApJS..122..207G} and MGPS-2
\citep[]{2007MNRAS.382..382M,2014PASA...31...42G}, the MOST Supernova
Remnant Catalogue \citep[MSC;][]{Whiteoak_1996}, and the Molonglo
Southern 4 Jy Survey \citep{2006AJ....131..100B} (the Southern
counterpart to the Northern 3CR catalog). It was particularly useful
for rapid follow-up observations of transient sources. For example, it
detected the prompt radio emission from supernova SN1987A
\citep{Turtle_1987} and discovered radio emission from the
micro-quasar GRO 1655-40 \citep{1994IAUC.6052....2D}. A systematic
monitoring campaign of SN 1987A led to its re-detection as a radio
supernova remnant \citep{Staveley_Smith_1992}.

\subsection{The SKAMP Upgrades}

In 2004, staged upgrades were proposed for the MOST as part of the
``Square Kilometre Array Molonglo Pathfinder'' (SKAMP) project, to
give the instrument its third major lease of scientific life. The
first upgrade, SKAMP-1 \citep{Adams_2004}, replaced the analogue
beamformer with an 8-bit digital correlator while retaining the same
bandwidth and field of view. The second upgrade, SKAMP-2
\citep{de_Souza_2007}, increased the system bandwidth from 3 to 30 MHz
and the instantaneous field of view by a factor of two (by processing
all 352 radio frequency outputs from the 352 independent outputs of
the telescope rather than pre-combining them into 176 pairs). This
involved laying over 200 km of optical fibre to 88 new 4-input digital
receivers (RX boxes) in the field.  Each RX box received its inputs
from one 18-m element and the digital optical transmission from the RX
boxes replaced the analogue coax cable used in MOST. Transmission of
clocks to the RX boxes was also upgraded to optical fibre. The SKAMP-2
optical fibre outputs (from the 88 RX boxes) feed 22 polyphase
filterbanks (PFBs) in the main control room. Each PFB is built on an
ATCA board using Xilinx Virtex 4 FPGAs (Field Programmable Gate
Arrays).

In late 2013, a new system design (the UTMOST) was proposed that could
add greater flexibility to the signal processing chain by using
software and commodity off-the-shelf servers and graphics processing
units instead of the Field Programmable Gate Arrays proposed for use
in SKAMP-2. Signals to the GPUs would be provided by the MOST
cylindrical reflector and feed, and the SKAMP-2 RX boxes and FPGA
filterbanks.

\subsection{Radio Astronomy in Software}

An emerging trend in radio astronomy over the past decade is the
replacement of custom chips with commodity, off-the-shelf (COTS)
signal processors such as Graphics Processing Units (GPUs) and
software. While less energy efficient than correlators using
Application Specific Integrated Circuits (ASICs) or Field Programmable
Gate Arrays (FPGAs), software correlators and signal processors are
particularly inexpensive to develop.  Changes to the configuration and
operational modes are easy to implement after construction, as this is
predominantly achieved in software. Examples of software correlators
(or ``coherent dedispersers'') currently in use for pulsar
observations are CPSR2 \citep{Bailes_2009} and GUPPI
\citep{DuPlain_2008}. Further, it has become straightforward to
publish and distribute software instruments such as the pulsar
processor \textsc{dspsr} \citep{van_Straten_2011} via online
repositories, for installation by other users. Other examples of
software baseband processing instruments are the VLBI correlator
\textsc{DiFX} \citep{Deller_2007} and \textsc{xGPU} \citep{Clark_2012}
a GPU-based correlator for ``large-N'' interferometers such as the
Murchison Widefield Array (MWA) \citep{Tingay_2013} and the
Large-Aperture Experiment to Detect the Dark Ages (LEDA)
\citep{2014era..conf10301G}.

A software backend for the MOST/SKAMP was proposed in early 2013, when
its potential to become a powerful Fast Radio Burst discovery engine
working commensally with other observing modes was realised. The
UTMOST retains the digital receivers and polyphase filterbanks of the
original SKAMP-2 system, but performs the rest of the signal
processing on commodity CPUs and GPUs. The UTMOST Software Correlator
has been designed to provide great processing flexibility and has
commensal modes that enable simultaneous mapping, pulsar coherent
dedispersion, burst-searching and baseband dump modes whilst employing
novel interference mitigation strategies.

This paper describes the scientific capability and design of the
UTMOST, the Swinburne University of Technology CPU/GPU backend
solution for the MOST. In \S 2 we provide a brief description of all
system components and their technical specifications. The site is
subject to interference from domestic mobile phones and communications
towers, and a simple technique to detect and attenuate their signals
is described in \S 3. Examples of the major observing modes are
provided in \S 4, the initial science programs are described in \S 5,
and our summary and future prospects are given in \S 6.

\section{UTMOST System Design}

\begin{figure*}
\begin{center}
\includegraphics[width=160mm]{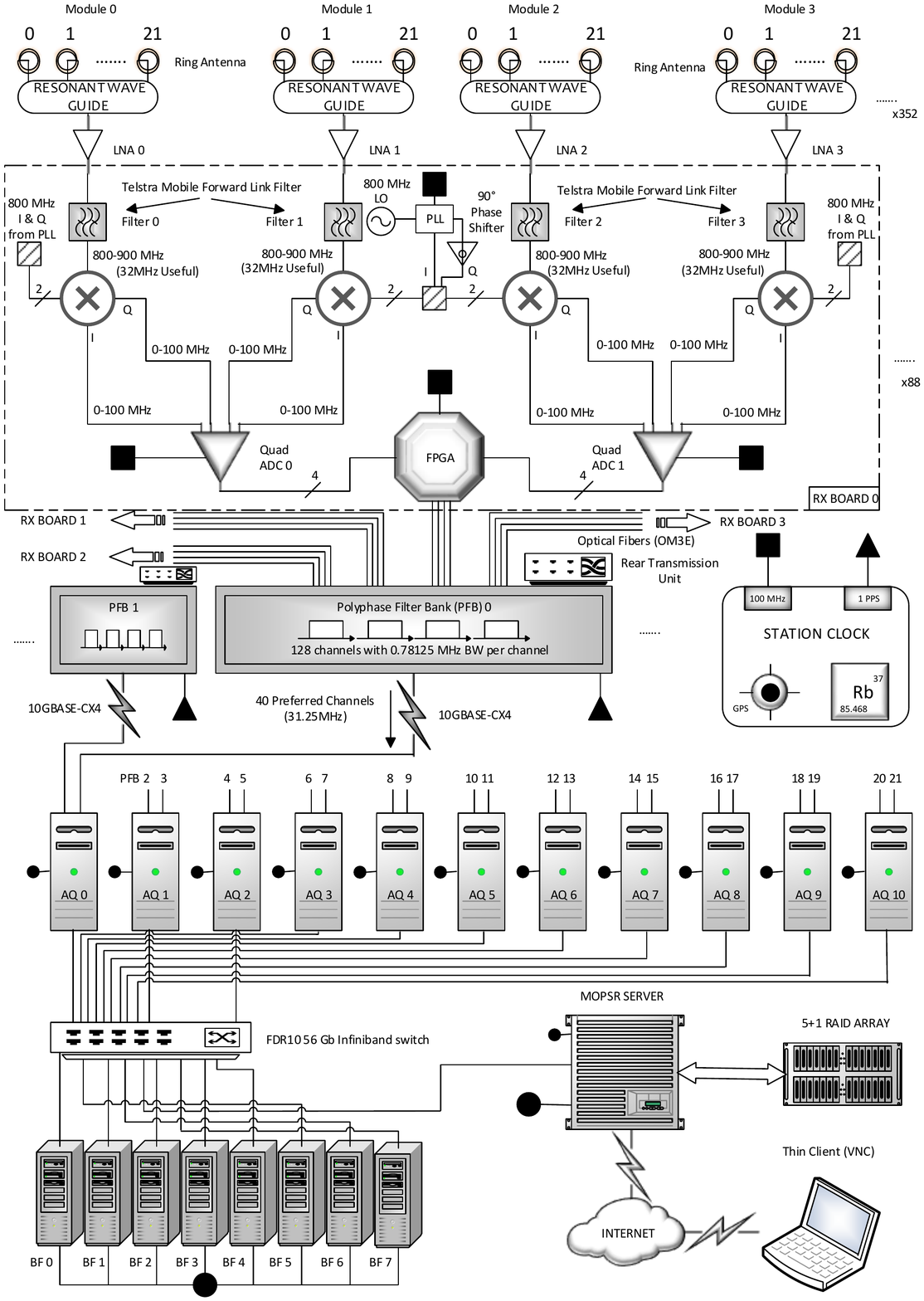}
\caption{The UTMOST block diagram. From the ring antennas to the LNAs
  the system is identical to that used on the MOST. The SKAMP-2
  upgrade includes the RX boxes and the polyphase filterbank (PFB), at
  which point the UTMOST backend taps off data via 10 Gb UDP packets
  using the CX-4 ports on the PFBs. The UTMOST uses commodity
  off-the-shelf computing hardware (marked ``AQ'' and ``BF'' nodes in
  the figure) and highly flexible software algorithms to achieve its
  signal processing requirements. AQ denotes the system's ``data
  acquisition'' nodes and BF the system's ``beam forming'' nodes.}
\label{UTMOSTblockdiagram}
\end{center}
\end{figure*}

The MOST consists of two parabolic cylinders 11.6\,m wide and 778\,m
long separated by a 15-m gap. Parabolic cylinders enable a linear
scaling of cost with collecting area but make the steering of the beam
along its axis rather complex. The MOST uses 352 modules each
comprising 22 ring antennas of circumference $\lambda$ (36 cm) for
operation at the central frequency of the 843 MHz feed. These ring
antennas are coherently summed in resonant cavities along the focal
line of the paraboloid, before being fed to 352 Low Noise Amplifiers
(LNAs) \citep{Robertson_1991}. The signals from each LNA are filtered
and down-converted before being 8-bit complex sampled in the field at
100 Megasamples/sec (i.e $200 \times 10^6$ B/s). This is equivalent to
Nyquist sampling a bandwidth of 100 MHz at a rate of 200 Msamples/sec.
They are transmitted via multicore multi-mode fibre to digital
Polyphase Filterbanks (PFBs) in a central building 100\,m north of the
junction between the two arms of the telescope. Here the data stream
is processed by the UTMOST software correlator backend.  A block
diagram of the system is shown in Figure \ref{UTMOSTblockdiagram} and
the basic telescope parameters provided in Tables 1 and 2. Each of
these stages is now described in more detail.

\begin{table}
\begin{center}
\label{utmosttable}
\caption{UTMOST telescope parameters.}
\begin{tabular}{l|c}
\toprule
Collecting Area & 18000 m$^2$\\
Number of RF inputs & 352\\
Operating band & 820-850 MHz\\ 
Primary beam dimensions & 4.25 deg $\times$ 2.8 deg\\
Fan beam dimensions & 46 arcsec $\times$ 2.8 deg\\
Theoretical gain & 3.5 K Jy$^{-1}$\\
Effective gain (Nov 2016) & $\sim 0.5$ K Jy$^{-1}$\\
SEFD (typical) & 115 Jy\\
$T_{\rm sys}$ (best inputs) & 100 K\\
\bottomrule
\end{tabular}
\end{center}
\end{table}

\begin{table*}
\begin{center}
\caption{Frequency resolution, time resolution, number of fan beams and
  coherent dedispersion properties for the UTMOST's operational modes.}
\begin{tabular}{l|lc}
\toprule
{\bf Single Module Mode} & Processed bandwidth & 31.25 MHz\\
& Number of frequency channels & 128\\
& Time resolution & 655.36 $\upmu$s\\
& Output data rate & 0.19 MB/sec/module\\
\hline
{\bf LORES Fan Beam Mode} & Number of frequency channels & 128\\
& Time resolution & 655.36 $\upmu$s\\
& Simultaneous pulsar coherent dedispersion & Yes\\
& Number of fan beams & 352\\
& Fan beam dimensions (FWHP) & 46 arcsec $\times$ 2.8 deg\\
& Output data rate & 0.19 MB/sec/module\\
\hline
{\bf HIRES Fan Beam Mode} & Number of frequency channels & 1024\\
& Time resolution & 327.68 $\upmu$s\\
& Simultaneous pulsar coherent dedispersion & Yes\\
& Number of fan beams & 704\\
& Fan beam dimensions (FWHP) & 46 arcsec $\times$ 2.8 deg\\
& Output data rate & 1.56 MB/sec/module\\
\hline
{\bf FX Calibration Mode}& Number of frequency channels & 1024\\
& Number of baseline products & $8 \times 351 = 2808$\\
& Simultaneous pulsar coherent dedispersion & No\\
\hline
{\bf FX Correlator Mode} & Number of frequency channels & 128\\
& Number of baseline products & ${352}\times{351}/2 = 61776$\\
& Simultaneous pulsar coherent dedispersion & No\\
\bottomrule
\end{tabular}
\end{center}
\end{table*}

\subsection{The Front-end Feeds}

In order to steer the primary beam, each telescope section in the
East-West direction uses an ingenious system of resonant ring antennas
which are sensitive to right-hand (IEEE) circularly polarised
radiation. The ring antennas are differentially rotated to steer the
phased-array beam and coherently add incident radiation in groups of
22 antennas spanning 4.57\,m of telescope length, producing 352
independent outputs called modules (176 on each arm). In the
North-South direction steering is comparatively simple, as each
telescope arm can be mechanically tilted from the zenith, enabling
observation of all parts of the sky south of declination $\delta =
+17^\circ$.

The gain of each module is approximately $G=0.01$\,K Jy$^{-1}$, where
we include an effective aperture factor $\eta = 0.5$. The system
temperature in the best performing modules is estimated, from
observations of point source calibrators, to be $T_{\mathrm sys} \sim
100$ K. The ratio $G/T_{\mathrm sys}$ presently can vary considerably
(by up to an order of magnitude) from module to module, as legacy
sources of self-generated RFI are still being identified and removed
in the ongoing overhaul of the entire system. Currently, in a typical
observation, the best modules achieve a system equivalent flux density
(SEFD) of $\sim$10 kJy, while a typical performance figure for the
whole instrument is SEFD $=$ 110 Jy, due to the combination of many
factors, including incorrect phasing, ring antenna misalignment,
self-induced RFI and cross-talk, RX box performance, and deformations
in the telescope structure. Despite this range in performance,
substantial verification science has been carried out during the
upgrade (see \S 5).

The LNAs contribute $\sim$20 K (as measured in the laboratory) to the
total system temperature, with the remainder coming from a variety of
sources including mesh leakage, spillover and cable feeds.  The
effective gain of each module varies in a complex manner as it is
steered East-West, a function termed the ``meridian distance gain
curve'' \citep{2000PASA...17...72G}. This arises endemically in
parabolic antennas because of multiple reflections along the cylinder
which attenuate the signal to give an effective gain of $<0.01$ K
Jy$^{-1}$ per module. When added coherently on the meridian, the 352
modules of the antenna should combine to produce a gain of 3.5 K
Jy$^{-1}$ and an SEFD of $\sim30$ Jy.

Four adjacent modules feed the independent RF signals from their LNAs
via coaxial cable to one of 88 RX boxes.  Prior to the RX boxes, the
UTMOST signal chain is identical to the MOST.  The MOST operated by
coherently combining adjacent modules, whereas SKAMP-2 retains the
independent inputs from each group of 22 ring antennas and brings back
the digitized signals on optical fibre.
  
\subsection{The Digital Receiver RX boxes and 3G Filters}

The 88 RX boxes apply a pair of commercially sourced analogue filters
to each of the four inputs in series to filter out local
telecommunications 3G and 4G up-links and produce four band-limited
signals, currently from 820 to 850 MHz. These are down-converted with
an 850 MHz local oscillator, and the complex signals are transmitted
via four optical fibres using SerDes format (achieved on a Lattice
FPGA) to the central processing building.

\subsection{The Polyphase Filterbanks}

Optical fibres carrying the digitised signal from sets of 4 RX boxes
at a time (i.e. 16 inputs) are aggregated in the central processing
building in a Rear Transmission Module (RTM), which transmits the data
to polyphase filterbanks (PFBs). The PFBs have a CX-4 port, which is
used to send 8k UDP ethernet packets to industry standard (copper) 10
Gb network interface cards (NICs). The PFBs use the SKAMP-2 hardware
with firmware modified to bypass the cross connect, ``corner-turn''
and fine filterbank operation.  Only the 128 coarse filterbanks
remain, providing data with a 781-kHz channel spacing.  The original
SKAMP-2 filterbanks were oversampled by a factor of 32/27, to reduce
signal loss at the coarse channel edges. Two extra modes are
implemented for UTMOST: a critically sampled 128-channel filterbank
and a pass-through mode.  Selection of the mode is programmable.  The
pass-through mode proved pivotal in debugging the more complex modes
during system implementation, but the default run mode is the
critically-sampled one. The over-sampled mode is reserved for future
use, and may help eliminate loss of sensitivity between frequency
channels, at the cost of a slightly increased data rate. The
filterbank modes use 12-tap polyphase filters and time-tag the data
using the station 1PPS clock (1 pulse per second) as a reference. The
data samples are requantised to 8-bit real, 8-bit
imaginary. Currently, in ``LORES'' mode, the system produces 128
channels, each 781 kHz wide, and 40 of these are passed to the data
acquisition servers. In ``HIRES'' mode, it produces 1024 channels,
each 97 kHz wide, 320 of which are passed through to the data
acquisition servers.  In critically-sampled mode, the data rate is 8
Gb/s per PFB.

To this stage in the system, the hardware is just the SKAMP-2
design. In UTMOST, the I/O FPGA has been re-programmed to send the
data to the servers via CX-4 ports using 10 Gb ethernet. The radio
astronomy signal processing problem then becomes solely a computing
one, and can be addressed in high-level programming languages.

\subsection {The AQ Servers}\label{aqser}

The data acquisition (AQ) servers are comprised of 11 COTS
server-class machines each with a 6-core Intel CPU, 64 GB of RAM, a
128 GB SSD drive, a dual-port 10\,Gb ethernet Network Interface Card
(NIC), a 56\,Gb Mellanox Infiniband card and a GTX 690 graphics card
(GPU).

The \textsc{psrdada} software
library\footnote{\url{psrdada.sourceforge.net}} is used to catch the
5120-byte UDP packets. The UDP packets are taken from the NIC and
placed in a ring buffer in the DDR3 memory of the server, from where
they are moved via the PCIe bus to the GTX 690 GPU's memory. A finite
impulse response filter (FIR), comprising 25 taps using Hamming window
coefficients, is applied in the GPU to compensate for fractional
sample delays (both systemic and geometric), and an analysis of the
spectral kurtosis of the incoming signal is applied to help detect
deviations from RFI-free statistics (see \S \ref{RFIexcision} on RFI).

Each coarse PFB is synchronised to the observatory reference clock,
which supplies a frequency standard and 1PPS signal. The coarse PFB
uses these references to synchronise headers within the UDP packet
timestamps to UTC time. The AQ servers parse the timestamps to insert
the data into the ring buffer. The FIR filter is updated every 16384
samples (or every 167.78 ms). The data are processed by the AQ GPUs in
blocks of 16384 samples, which sets the update rate. This is a
convenient size for efficient data processing, while allowing us to
exceed the update rate required to ensure appropriately delayed
voltages for coherent beam formation.

Instrumental delays, phase offsets and weights are retained for each
antenna from the most recent calibration solution, which is typically
performed with bright, unresolved, radio sources (generally brighter
than 10 Jy). For calibration, eight modules are chosen from the array
as reference antennas. These modules are chosen to be well-functioning
high SNR units and are spread roughly uniformly along the array.  The
eight reference antennas are cross-correlated with all 352 modules,
forming 8 $\times$ 352 integrated cross-power spectra. The inverse
Fourier transform of the cross-power spectra yields the ``lag
function'', from which the instrumental delays and phase offsets are
computed relative to an (arbitrary) reference antenna. Triplets of
antenna pairs, on longer baselines ($>50$ meters) are used to compute
the closure delay and phase to reduce impact of local noise and RFI
affecting the calibration solution. The weighting scheme for each
antenna, used during tied-array beam forming, is computed as the
inverse of its measured S/N, normalised to the array maximum. These
delays and weights are applied by the AQ GPU pipeline, subsequent to
applying geometric delays for the calibration source's (known)
position. Each server handles a total data rate of 16 Gb/s.

Suspect data, as flagged by a measure of the spectral kurtosis, are
replaced by Gaussian random noise of the same power as the input
signal. This eliminates the worst of the 3G/4G phone
transmissions. The fraction of flagged data as a function of time and
frequency is retained so that down-stream data products can be
adjusted if necessary for the impact on the system sensitivity. In the
``single module" mode of the instrument it is possible to write out
filterbank data to disk for subsequent processing. These data can be
folded at the topocentric pulse period of a pulsar to test that any
given module is working prior to any correlation attempts. We also
search the known mobile phone handset transmission bands within our
820-850 MHz bandwidth for excess power, using a threshold determined
from the binomial distribution and the number of frequency channels
being monitored \citep{Nita_Gary2010}. In high frequency resolution
mode (HIRES), the phone bands span a sufficiently large number of
channels to make this an effective strategy: it is not used in our low
frequency resolution mode (LORES) as the number of channels is too
low.

\subsection{Frequency multiplexing}

Radio interferometers combine the same frequency channels from
different telescopes to form interference fringes. This is a
high-bandwidth, computationally-demanding task. In a standard hardware
correlator (eg \citep{2007A&A...462..801E}) this is usually done by
constructing identical correlator ``blocks'', each of which process
some fraction of the total bandwidth after passage through a
``corner-turn'' device. This device takes a series of time-major order
data and brings together frequency-major order data for
cross-correlation. In the UTMOST system this task is performed on the
servers using remote direct memory access (RDMA) transfers via FDR10
56 Gb infiniband using VMA Mellanox infiniband libraries. RDMA has the
advantage that it does not impact on the CPUs during transfers. The
choice of RDMA chunk size is driven by the GPU memory capacity on the
AQ servers. Larger chunk sizes are more efficient. For LORES mode,
this is 98,304 samples, and for HIRES mode it is 16,384 samples.

Post-cornerturn data are sent to the so-called ``BF'' servers
(described in detail in the next section) for both correlation and
beam-forming modes. The major advantage of a software correlator with
its appropriate ring buffers is that we can eliminate the need to
achieve the challenging and time-consuming process of synchronization
in the Field Programmable Gate Arrays.

\subsection{Beam former (BF) servers}\label{BFservers}

After the corner-turn, the data are aligned in time in the RAM of the
beam-forming servers, and the same frequency channels from every
module are co-located. At this point, the servers can perform one of
five operations: (i) detect and sum each voltage to form a single
\textit{incoherent} primary beam; (ii) add the complex voltages to
form \textit{coherent} tied-array beams (i.e. ``fan beams''); (iii)
cross correlate each input against the others (an ``X'' mode) to form
interference fringes; (iv) produce fine channels that are themselves
correlated for a limited subset of baselines (an ``FX'' spectral-line
mode); and (v) create multiple fan-beams on the sky. The technical
details of these operational modes are shown in Table 2 and discussed
fully in \S 4.

The BF servers consist of 8 COTS machines consisting of an 8-core
Intel CPU, 4 ``Titan X'' GPUs, 128 GB of RAM, a 440 GB SSD drive and a
single-port FDR 56 Gb ethernet (NIC), and were installed in May 2015. 

In the first operational mode, a power time series from the incoherent
primary beam is written to a filterbank file, and is invaluable for
RFI monitoring. In the other modes, voltages from a tied-array beam
can be detected and written to a filterbank file for offline
processing, or incoherently or coherently dedispersed and folded into
a pulsar profile. Coherent dedispersion is performed upon the voltage
tied-array beam using \textsc{dspsr} \citep{van_Straten_2011}.

Pulsar folding and timing with a tied-array beam requires only minor
compute resources, as the workload is distributed (in frequency)
across the BF servers. These tasks are easily performed on
CPUs. Pulsar timing is described in detail in \S \ref{pulsartiming}.

Map making has been performed from fringes using standard packages
such as Miriad and \textsc{CASA} (see Section \ref{fxcacm}). Maps have
also been made from the time series data produced in filterbank format
from the multiple fan beams but this is at an early stage of
development (see \S \ref{sfbm}).
  
\subsection{Real time data quality control}

A software correlator like the UTMOST provides many opportunities for
inspection of data and quality control in real time. For example, the
Lattice FPGA on the RX boards can send data snippets (of a few kB) via
Gb ethernet to the AQ machines, which can be plotted in real time as
system bandpasses, time series and data histograms on the observer's
web interface. Once per second the UDP capture system sends 512
voltage measurements per channel to a ring buffer from which a range
of plots can be produced, again available for scrutiny via a web
interface.

At the other extreme, we are able to record long (10 to 120 s real
time) bursts of the PFB raw voltages to the SSD drives of the AQ
computers for subsequent offline processing. The software backend can
either operate in real time on the real voltage streams or in an
offline mode on the recorded voltages. The usefulness of this mode in
debugging the software cannot be understated. A real-time correlator
cannot usually play data back, and thus cannot perform ``difference''
experiments or confirm the validity of new features on proven input
data sets. During the commissioning phases of the UTMOST correlator,
raw data captured in this way proved invaluable in debugging the
system: it could be played back an arbitrary number of times.

\section{Radio Frequency Interference (RFI) Excision}  
\label{RFIexcision}

Detecting, excising and/or avoiding RFI has always been a major issue
in radio astronomy and is becoming increasingly pervasive. UTMOST
operates in a part of the spectrum used by two Australian mobile phone
companies, and we often detect RFI from handsets, emitting typically
in bandwidths of $\sim$4 MHz, with greatly varying strength.

Tests early in the upgrade showed that it is rare that two or more
4-MHz handset bands are detectable at any given time, but that for
about 10\% of the time RFI from a single handset can be seen, most
typically during the morning, late afternoon and
evening. Nevertheless, handset emission can be detected at any time,
even during the much quieter hours from midnight to dawn. Handsets
typically emit pulses of $\sim$20 ms duration -- these are used to
register with the out-of-band base station towers operated by the
phone companies. The nearest base stations in frequency to our
operating bands are the 3G service at 890 MHz and the 4G service at
800 MHz. The ability to excise this highly prevalent RFI was crucial,
or the science would be seriously compromised.

As with most radio observations, our system is noise-dominated, and
good data are expected to be statistically well described by Gaussian
random noise. Mobile phone transmissions lead to non-Gaussian
distortions in the voltages and can be recognised by measuring the
spectral kurtosis of the voltage stream to search for sudden
deviations in the total power \citep{2010MNRAS.406L..60N}. Since
single modules operating at the target system temperature of $\sim$100
K have an SEFD of about 10 kJy, it is extremely unlikely that genuine
celestial signals will be deleted erroneously if we eliminate data
with a significant spectral kurtosis statistic.

\begin{figure}
\begin{center}
\includegraphics[height=0.8\columnwidth, angle=0]{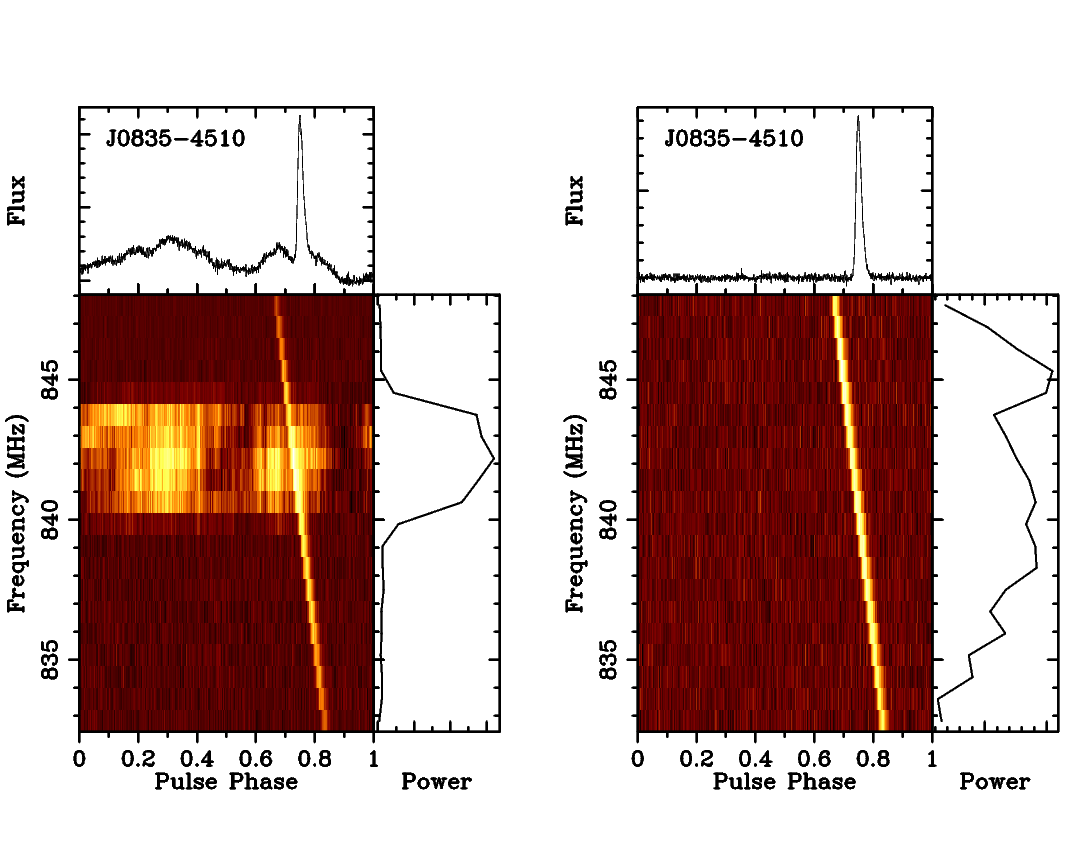}
\caption{Data taken on a single module of the Vela pulsar with a phone
  call occurring during the observation. In the left panel the data
  were processed without RFI rejection, while in the right panel the
  same data were processed using standard UTMOST RFI rejection
  procedures.}
\label{Vela_sk}
\end{center}
\end{figure}

In Figure \ref{Vela_sk} we show the pulsar profile from an observation
of the Vela pulsar where the raw voltages were written to disk and
processed in two ways.  In the first panel our ``standard'' RFI
rejection procedure was enacted. On a module by module basis the
spectral kurtosis of the input voltages were assessed for deviations
from Gaussianity. If above a threshold, the voltages were replaced by
random noise generated in the GPU with the same standard deviation and
mean as the uncorrupted data.  The data were then coherently summed to
form the pulsar profile. In the second panel, all of the input data
were similarly summed and processed but without any RFI rejection
algorithms. The UTMOST thus offers two levels of protection against
RFI. Large impulsive bursts are removed on a module by module basis,
and the weaker interfering signals are dephased by the addition of the
352 input voltages.

Our kurtosis and power-based detection methods currently search for
interference on timescales from 1--20 ms. The procedure operates live
on 40 coarse frequency channels on the 352 modules of the telescope,
but nevertheless only requires a fraction of the available GPU cycles
on the GTX 690s, which are mainly used for delay computations.

This RFI rejection methodology is used when calibrating the
instrument. Phone calls from certain distances/angles can still badly
affect some segments of the data, but on the whole, calibration is
still possible almost 100\% of the time.

It might be assumed that all dispersed pulses detected with an
interferometer were celestial, but it is clear that over limited
bandwidths (typically the phone bands) some combinations of angle,
distance and mobile phone velocity produced dispersed pulses in some
of the fan-beams. These usually appear in a large number of fan-beams
or rapidly traverse the fan-beams making them reasonably easy to
separate from far-field signals (see \S \ref{frbsearch}).

Research is underway to determine if spatial filtering techniques
\citep[e.g.][]{1998AJ....116.2598B} can be used to subtract the phone
transmissions, possibly in real-time, and to determine whether such
sources of RFI can be used to phase calibrate.

\section{Observing Modes}

In this section we describe the major science modes in which UTMOST
can currently operate. All of these modes are defined in software and
trivial to switch between : such flexibility of the software has been
key to the rapid commissioning.
 
\subsection{Single Module Mode}


In single module mode, bright pulsars can be detected by incoherently
summing folded profiles from each of the 352 modules of the
telescope. With the advent of our capability to form a tied-array beam
using an arbitrary number of modules (see ``Tied-Array Beam Mode''
below), incoherent timing mode was effectively superseded. We
nevertheless retain the capability to observe in this mode, as it is a
quick method of characterising the response of each module
independently of phased-array modes. This has proved very useful
during the commissioning phases, for example, as a means of
identifying poorly performing modules. Table 2 shows the technical
details about this and further modes described below.

\subsection{Frequency Multiplexed Modes}


We have been able to produce a tied-array beam using all 352 antennas
and 40 frequency channels since the installation of the BF servers in
May 2015 (\S \ref{BFservers}). Tied-array beams can be formed at
an arbitrary sky position to track a source, or it can be set to an
arbitrary (fixed) position and allow objects to drift through the
beam. Pulsars can also be folded live, according to their current
ephemerides, while operating the tied-array beam (cf. \S 5.1). To
date more than 300 pulsars have been detected and timed in this mode,
and more than 50 have been seen with individual pulse detections. We
almost always operate this mode concurrently with fan beam mode, to
search for single pulses over the entire primary beam.

\subsubsection{LORES Fan Beam Mode}\label{sfbm}

\begin{figure}
\begin{center}
\includegraphics[width=70mm,angle=-90]{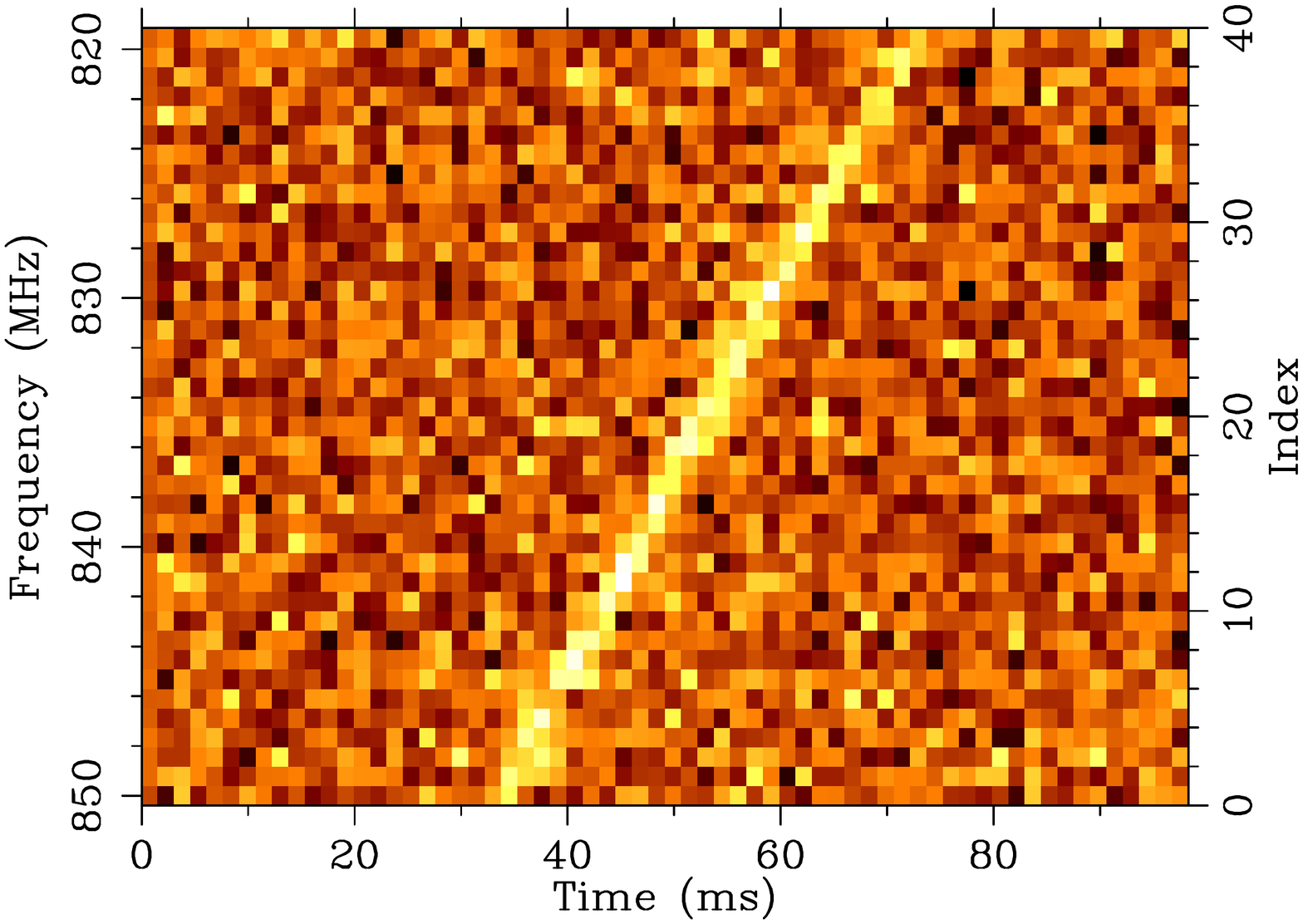}
\caption{
A single pulse from the pulsar PSR J1745--3040 detected in fan-beam
mode at a trial width of 10 ms with a detection significance of 17 sigma.
The trial dispersion measure was within 0.1 pc cm$^{-3}$ of the pulsar's
catalogued value of 88.5 pc cm$^{-3}$. 
}
\label{single_pulse}
\end{center}
\end{figure}

Tiling of the primary beam is achieved with a set of 352 narrow ``fan
beams'', arranged uniformly and parallel to the meridian. These are
generated by simply rotating the complex voltages to phase on to a set
of locations on the sky, and as such are just instances of 352
``tied-array beams''. Their spacing can be arbitrarily set, and is
typically set up to overlap near their half-power points ($\approx
46''$). In the direction parallel to the meridian, fan beams have the
same extent as the primary beam (i.e. full-width half-power of 2.8
degrees). Twice this number of fan-beams has been trialled
successfully, so that options exist to tile the primary beam more
densely, or to cover a larger area of sky. More fan beams still would
require additional hardware capacity.

Fan beam mode has been extensively tested with pulsars, by allowing
them to transit the primary beam, and was used in 2015 and 2016 to
search for FRBs and single pulses from pulsars (\S \ref{frbsearch}). In
Figure \ref{single_pulse} a single pulse is shown that was detected in
our pipeline from the bright pulsar PSR J1745--3040.  Fan-beam mode
can also be used to construct maps of the sky, in the same manner used
for the SUMSS survey. A pipeline for the construction of maps from
fan-beam mode data remains at an early stage of development. From
November 2016, we have also been able to form tied-array beams and
fold on and time up to four pulsars in the primary beam area, fully
commensally with fan beam mode.

\subsubsection{HIRES Fan Beam Mode}

A high frequency resolution and higher spatial resolution version of
the fan beam mode has been available since November 2016. This
``HIRES'' mode has 8 times the frequency resolution, so that the
effects of dispersion smearing are substantially reduced (the DM
smearing is now only $\sim 1.2$ ms for DM $= 1000$ pc\,cm$^{-3}$). Our
three FRBs \citep{2017MNRAS.468.3746C} were found in the lower
resolution mode prior to implementing this system. It has been the
standard operational mode for FRB searches and pulsar timing since
November 2016.

\subsubsection{FX Calibration and Correlation Modes}\label{fxcacm}

The UTMOST has two main software correlator modes. In the FX mode, an
FFT further channelises the input data to approx 24 kHz wide, then 8
reference modules are cross-multiplied with all 352 modules
(cf. \S\ref{aqser}).  This is primarily used for phasing the array
when the calibration is very uncertain. In the X mode all 352 modules
are cross correlated with each other, forming 352$\times 351/2$
baselines, but at the coarse frequency resolution of the PFBs. This
mode uses Clark's \textsc{xGPU} code \citep{Clark_2012}.

The output data products from the FX and X modes are averaged in time,
usually between 20 and 60 seconds. The different numbers of output
channels in FX and X modes also influence the total data rate, which
on average is 10.0 MB/sec.

The number of floating-point operations required to cross correlate
352 antennas is 8 FLOPS $\times$ 31.25 MHz $\times$ 67716 baselines,
or 15.44 Tflops. There are 32 TitanX GPUs in the BF nodes, such that
spreading our 31.25 MHz across them means each one is only required to
do 0.976 MHz of bandwidth, or perform at about 0.486 TFLOPS. Our GPUs
have a theoretical performance of 6.144 TFLOPS, so there is ample
headroom. We note that, in practice, a GPU code can rarely run at the
theoretical peak performance, as real-world constraints impose
themselves. Tests show that on our BF modes, where we get
\textsc{xGPU} to process as much bandwidth as possible on a single
GPU, TitanXs can process about 6.86 MHz each, at 3.39 TFLOPs, or at
approximately 55\% of the theoretical maximum of 6.144 TFLOPS.  In
daily operations, each BF server processes 3.9 MHz of bandwidth, with
each GPU processing 0.976 MHz of bandwidth.

At our operating frequency of 843 MHz, and with a 1.6-km maximum
baseline length, the bandwidth decorrelation for the 700-kHz channels
is less than 1\% over the 2.8 degrees FWHP of the primary beam.

\begin{figure}
\begin{center}
\includegraphics[width=80mm]{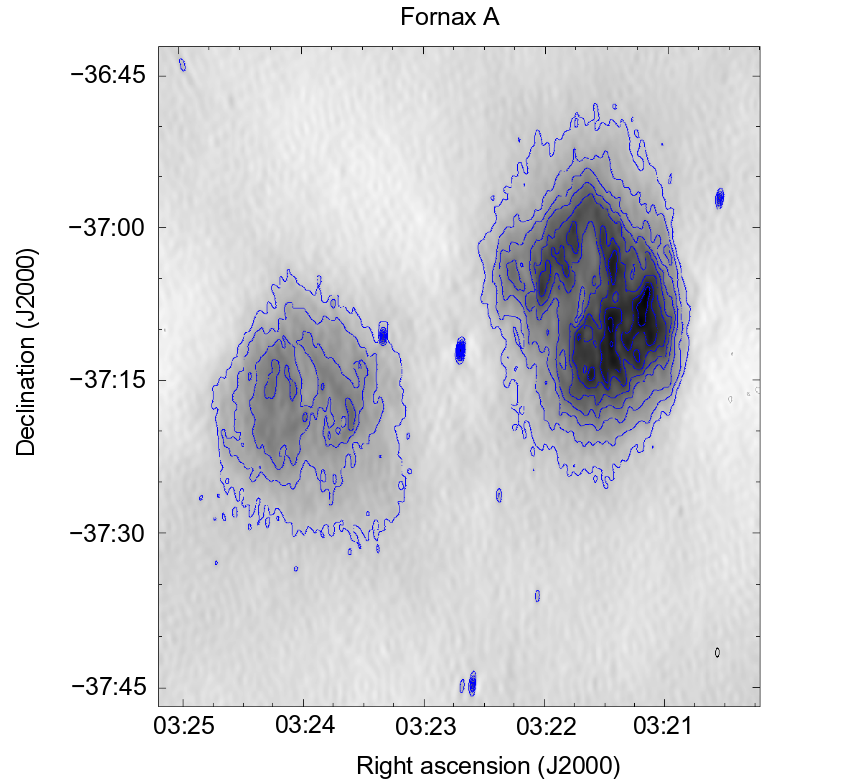}
\caption{Fornax A, from a 9-hour observation of the source made in
  November 2014. Only 25\% of the telescope collecting area and a
  small fraction of the bandwidth were used to produce this image. A
  nearby phase calibrator was observed every 2 hours. The image noise
  is $\sim$5 mJy/beam and is dominated by remaining artifacts rather
  than thermal noise. The contour levels are at $-$25, 25, 50, 75
  {\ldots} mJy/beam. Short spacings were deleted to improve the image
  quality as these were compromised by RFI.}
\label{FornaxA}
\end{center}
\end{figure}

In Figure \ref{FornaxA} we show the first synthesis map made with the
UTMOST, showing a 1 deg$^1$ field centred on the bright Southern radio
source Fornax A. The data were taken in November 2014 and the image is
compromised by the need to delete the short spacings ($\lesssim100$ m) and
some frequencies due to RFI. Only 25\% of the telescope's collecting
area was used. Significant improvements have since been achieved and
we can now correlate and take data in such modes from all 352 modules.

\section{Major Science Goals and System Performance}

A major scientific driver for the UTMOST project is to discover Fast
Radio Bursts (FRBs), but as we will demonstrate, this does not need to
come at the cost of other scientific objectives. Most, if not all of
the science modes being implemented, will eventually operate
commensally with our FRB search modes.

\subsection{Pulsar Timing}\label{pulsartiming}

At present the UTMOST typically times a single pulsar in the primary
beam of the telescope, but we can time up to four pulsars at once.  In
a 24-hr run the UTMOST can obtain useful pulsar arrival times from
$\sim$100 sources and in a complete week, $\sim$ 300 individual
sources.

\begin{figure}
\begin{center}
\includegraphics[width=80mm]{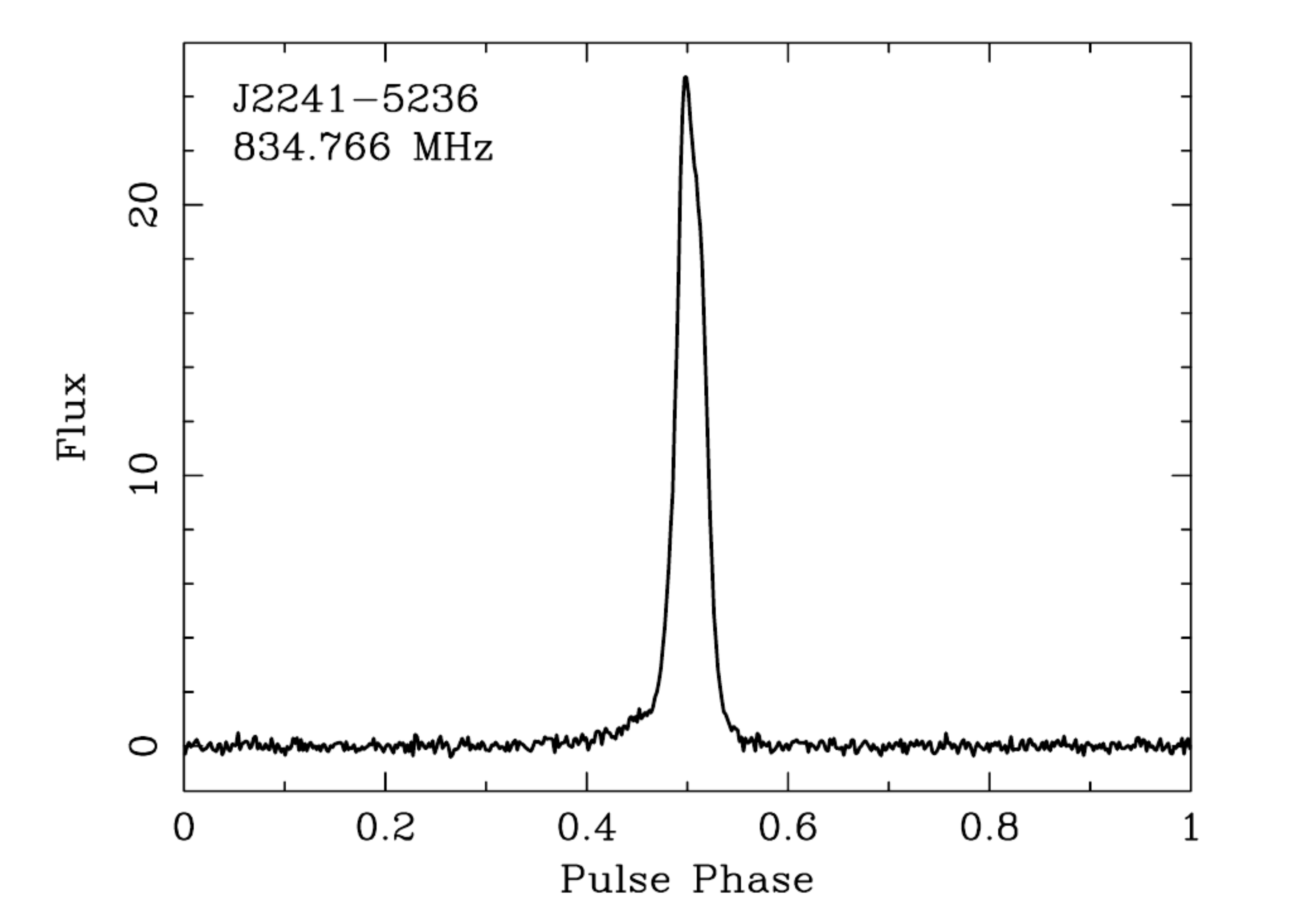}
\caption{Pulse profile for the millisecond pulsar PSR J2241--5236 in
  a 60 minute observation with the UTMOST telescope.  The bandwidth is
  31.25 MHz and the pulsar has been coherently dedispersed.  }
\label{2241profile}
\end{center}
\end{figure}

\begin{figure}
\begin{centering}
\includegraphics[width=70mm,angle=-90]{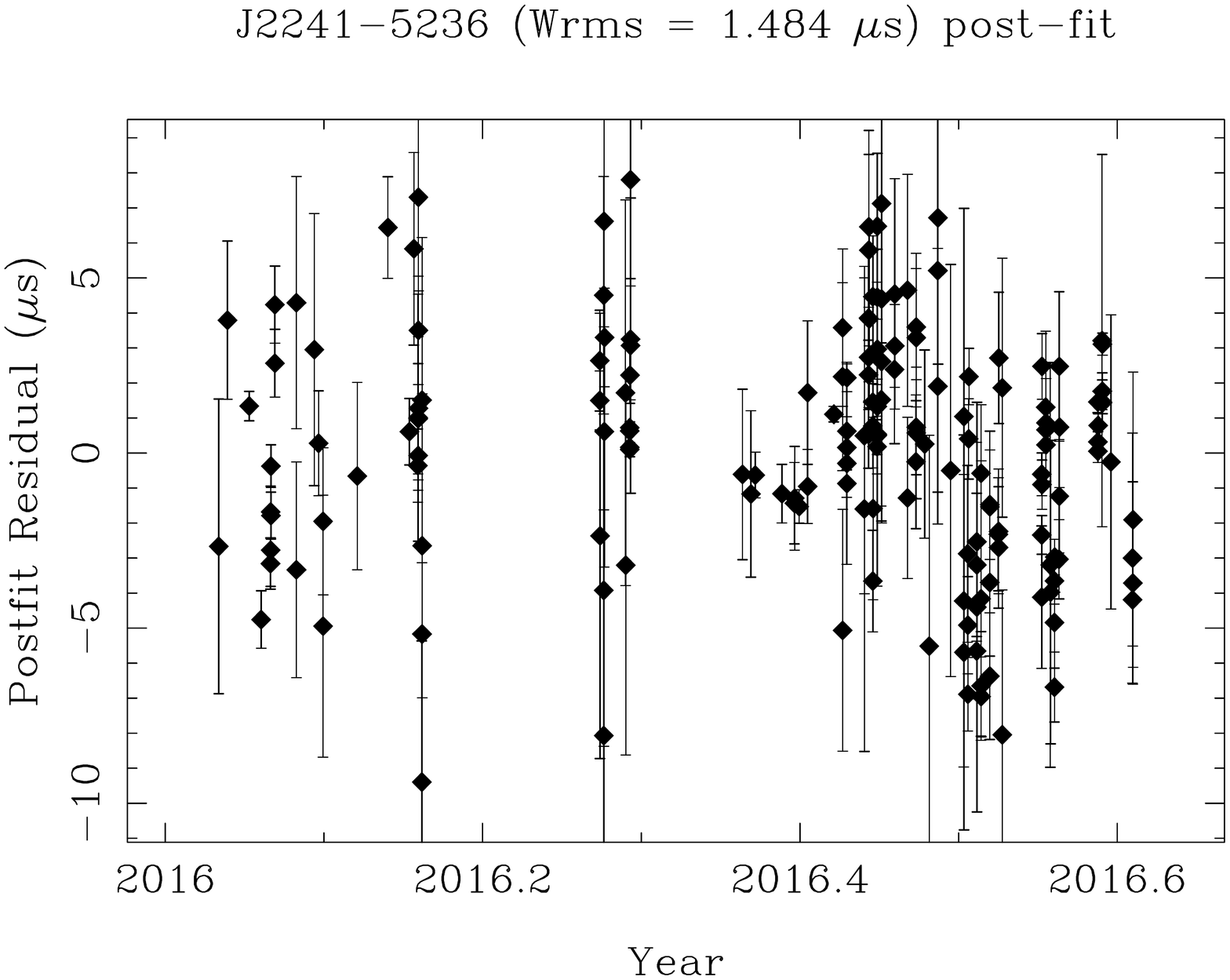}
\caption{Pulse timing residuals for the binary millisecond pulsar PSR
  J2241--5236 observed with the UTMOST.  Despite the relatively poor
  effective system temperature, sub-microsecond residuals are not
  uncommon on this pulsar whose flux density varies considerably due
  to interstellar scintillation.  }
\label{2241timing}
\end{centering}
\end{figure}

Detecting only right-hand circular polarisation limits the accuracy of
arrival times on pulsars with complex and/or highly polarised
profiles, as their profile is a function of meridian angle. On the
bright and highly-polarised millisecond pulsar PSR J0437--4715 timing
residuals are clearly divergent at large hour angles, but nevertheless
accurate to $\sim$5 $\upmu$s. The best pulsar for timing is the bright
millisecond pulsar PSR J2241--5236, whose narrow pulse profile (see
Figure \ref{2241profile}) has permitted $\sim$2 $\upmu$s rms residuals
(Figure \ref{2241timing}). Our pulsar timing program will be reported
in Jankowski et. al (in prep).

During the commissioning of UTMOST, 7 glitches in pulsar pulse arrival
times have been discovered \citep{2015ATel.6903....1J,
  2015ATel.8298....1J, Jankowski2015c}. This is an excellent
illustration of the power of timing on near daily cadence at a
facility like UTMOST.

A key achievement of the system is that both the source selection and
the duration of the observation are largely automatic, governed by a
series of Python scripts -- the ``Dynamic Scheduler''. For the fully
commissioned system, the goal is to time several hundred pulsars per
day. The Dynamic Scheduling system, commissioned in June 2016, has
increased the observing efficiency substantially, in terms of total
numbers of pulsars observed to a particular accuracy. The scheduler
can decide, on the basis of the accuracy of the timing measurement of
the pulsar, or its signal-to-noise ratio, when to terminate the
observation and move on to the next pulsar (lists of which are
requested as pulsar observations near completion). Substantial
increases in observing efficiency are anticipated as this system
matures.

\subsection{FRB search program}\label{frbsearch}

Fast Radio Bursts (FRBs) are radio transients which are bright
($\sim$1 to 100 Jy ms), have high dispersion measures (300 to 2500
pc\,cm$^{-3}$) and short durations (of order a few ms) and a high sky
rate of $\approx 2.5 \times 10^3$ events/sky/day above a fluence of
$\approx$ 2 Jy\,ms at 1.4 GHz \citep{KeanePetroff}. Only 25 bursts
have been published to date, most of which have been found in pulsar
search programs operating at the Parkes radio telescope over the last
decade. FRBs have considerable potential for probing the properties of
the ionised Intergalactic Medium, and are of course highly intriguing
in their own right \citep{JPska}. An FRB has been found to repeat
\citep{2016Natur.531..202S} and has been subsequently associated with
a dwarf galaxy \citep{Tendulkar_2017} at a redshift of $z \approx
0.2$.

UTMOST is operating an ongoing FRB search program concurrently with
other science modes (typically during pulsar timing, but also
independently when the telescope is parked, similarly to the SUPERB
project being operated at Parkes \citep{Superb1}). Our search methods
and validation tests, and our first two search programs for FRBs (at
low sensitivity due to the ongoing commissioning) are described in
\cite{2016arXiv160102444C}.

A third survey, carried out from February to November 2016, yielded 3
FRBs \citep{2017MNRAS.468.3746C}, while the system was at only about
15\% of its theoretical sensitivity. The detection of FRBs while
commissioning the system is very encouraging for the future of the FRB
search program. If all modules performed at their optimal (SEFD of 10
kJy), we might expect to detect an FRB every week or so. This would
constrain the spectral index distribution of the events and their
cosmological distribution \citep[e.g.][]{2016MNRAS.458..718C}.

\section{Summary and Prospects}

We have described the UTMOST, a new powerful backend for the Molonglo
radio telescope.  It is a hybrid solution to signal processing
requirements, replacing a frequency multiplexer, a correlator and fine
filterbanks of the SKAMP-2 system, with software and commodity
off-the-shelf hardware. The solution was inexpensive, rapidly deployed
and highly flexible compared to traditional approaches. In many
senses, the telescope is robotic, using software to select both
sources and the duration of the observation, based upon predefined
rules and near real-time feedback from the data.

The telescope now has twice the field of view and ten times the
bandwidth of the original MOST. The FX software correlator is superior
to the MOST's single-channel fan-beam multiplication interferometer
(designed in the 1970s), and includes novel interference rejection
algorithms.

Inherent to the solution is the capacity to run commensal
modes. UTMOST can simultaneously make images, coherently dedisperse
pulsars, perform real-time searches of coherent fan-beams for
dispersed single pulses out to dispersion measures of several 1000 pc
cm$^{-3}$, excise radio frequency interference in real time and offer
a range of new diagnostic information on system performance. The
record and playback capability facilitates rapid debugging of new
modes. Pulsar timing and FRB searches have commenced, as well as
searches for rotating radio transients and 1- and 2-D synthesis
imaging.

\section*{Acknowledgements}

We acknowledge the Australian Research Council grants CE110001020
(CAASTRO) and the Laureate Fellowship FL150100148.  The Molonglo
Observatory is owned and operated by the University of Sydney with
support from the School of Physics and the University. We thank the
referee for many insightful comments and suggestions. We acknowledge
many varied contributions and advice from Jay Banyer, Mike Kesteven,
Ron Koenig, Tom Landecker, Greg Madsen, Joseph Pathikulangara, Kathryn
Plant, Ludi de Souza, Darshan Thakkar, Glen Torr, Jamie Tsai, John
Tuthill and Ding Yan. The late Professor George Collins allocated
strategic funds for the purchase of the supercomputer in use at the
facility from Swinburne University and was a passionate advocate for
the project. The CSIRO Astronomy and Space Sciences division provided
support for modifications to the SKAMP-2 design.

\bibliographystyle{mnras}

\nocite*{}
\bibliographystyle{pasa-mnras}
\bibliography{utmost}

\begin{thebibliography}{}
\makeatletter
\relax
\def\mn@urlcharsother{\let\do\@makeother \do\$\do\&\do\#\do\^\do\_\do\%\do\~}
\def\mn@doi{\begingroup\mn@urlcharsother \@ifnextchar [ {\mn@doi@}
  {\mn@doi@[]}}
\def\mn@doi@[#1]#2{\def\@tempa{#1}\ifx\@tempa\@empty \href
  {http://dx.doi.org/#2} {doi:#2}\else \href {http://dx.doi.org/#2} {#1}\fi
  \endgroup}
\def\mn@eprint#1#2{\mn@eprint@#1:#2::\@nil}
\def\mn@eprint@arXiv#1{\href {http://arxiv.org/abs/#1} {{\tt arXiv:#1}}}
\def\mn@eprint@dblp#1{\href {http://dblp.uni-trier.de/rec/bibtex/#1.xml}
  {dblp:#1}}
\def\mn@eprint@#1:#2:#3:#4\@nil{\def\@tempa {#1}\def\@tempb {#2}\def\@tempc
  {#3}\ifx \@tempc \@empty \let \@tempc \@tempb \let \@tempb \@tempa \fi \ifx
  \@tempb \@empty \def\@tempb {arXiv}\fi \@ifundefined
  {mn@eprint@\@tempb}{\@tempb:\@tempc}{\expandafter \expandafter \csname
  mn@eprint@\@tempb\endcsname \expandafter{\@tempc}}}

\bibitem[\protect\citeauthoryear{Adams, Bunton  \& Kesteven}{Adams
  et~al.}{2004}]{Adams_2004}
Adams T.~J.,  Bunton J.~D.,   Kesteven M.~J.,  2004, \mn@doi [Exp Astron]
  {10.1007/s10686-005-2861-y}, 17, 279

\bibitem[\protect\citeauthoryear{Amy, Hunstead  \& Vaughan}{Amy
  et~al.}{1989}]{Amy_1989}
Amy S.,  Hunstead R.,   Vaughan A.,  1989, PASA, 8, 172

\bibitem[\protect\citeauthoryear{Bailes}{Bailes}{2009}]{Bailes_2009}
Bailes M.,  2009, \mn@doi [Proceedings of the International Astronomical Union]
  {10.1017/s1743921309990421}, 5, 212

\bibitem[\protect\citeauthoryear{{Barnbaum} \& {Bradley}}{{Barnbaum} \&
  {Bradley}}{1998}]{1998AJ....116.2598B}
{Barnbaum} C.,  {Bradley} R.~F.,  1998, \mn@doi [\aj] {10.1086/300604}, \href
  {http://adsabs.harvard.edu/abs/1998AJ....116.2598B} {116, 2598}

\bibitem[\protect\citeauthoryear{{Bock}, {Large}  \& {Sadler}}{{Bock}
  et~al.}{1999a}]{1999AJ....117.1578B}
{Bock} D.~C.-J.,  {Large} M.~I.,   {Sadler} E.~M.,  1999a, \mn@doi [\aj]
  {10.1086/300786}, \href {http://esoads.eso.org/abs/1999AJ....117.1578B} {117,
  1578}

\bibitem[\protect\citeauthoryear{Bock, Large  \& Sadler}{Bock
  et~al.}{1999b}]{Bock_1999}
Bock D. C.-J.,  Large M.~I.,   Sadler E.~M.,  1999b, \mn@doi [The Astronomical
  Journal] {10.1086/300786}, 117, 1578

\bibitem[\protect\citeauthoryear{{Burgess} \& {Hunstead}}{{Burgess} \&
  {Hunstead}}{2006}]{2006AJ....131..100B}
{Burgess} A.~M.,  {Hunstead} R.~W.,  2006, \mn@doi [\aj] {10.1086/498677},
  \href {http://adsabs.harvard.edu/abs/2006AJ....131..100B} {131, 100}

\bibitem[\protect\citeauthoryear{Burke-Spolaor, Bailes, Ekers, Macquart  \&
  Crawford}{Burke-Spolaor et~al.}{2011}]{Burke_Spolaor_2011}
Burke-Spolaor S.,  Bailes M.,  Ekers R.,  Macquart J.-P.,   Crawford F.,  2011,
  \mn@doi [{ApJ}] {10.1088/0004-637x/727/1/18}, 727, 18

\bibitem[\protect\citeauthoryear{{Caleb}, {Flynn}, {Bailes}, {Barr},
  {Hunstead}, {Keane}, {Ravi}  \& {van Straten}}{{Caleb}
  et~al.}{2015}]{2015arXiv151202738C}
{Caleb} M.,  {Flynn} C.,  {Bailes} M.,  {Barr} E.~D.,  {Hunstead} R.~W.,
  {Keane} E.~F.,  {Ravi} V.,   {van Straten} W.,  2015, preprint, \href
  {http://adsabs.harvard.edu/abs/2015arXiv151202738C} {} (\mn@eprint {arXiv}
  {1512.02738})

\bibitem[\protect\citeauthoryear{{Caleb} et~al.,}{{Caleb}
  et~al.}{2016a}]{2016arXiv160102444C}
{Caleb} M.,  et~al., 2016a, preprint, \href
  {http://adsabs.harvard.edu/abs/2016arXiv160102444C} {} (\mn@eprint {arXiv}
  {1601.02444})

\bibitem[\protect\citeauthoryear{{Caleb} et~al.,}{{Caleb}
  et~al.}{2016b}]{2016MNRAS.458..718C}
{Caleb} M.,  et~al., 2016b, \mn@doi [\mnras] {10.1093/mnras/stw109}, \href
  {http://esoads.eso.org/abs/2016MNRAS.458..718C} {458, 718}

\bibitem[\protect\citeauthoryear{{Caleb} et~al.,}{{Caleb}
  et~al.}{2017}]{2017MNRAS.468.3746C}
{Caleb} M.,  et~al., 2017, \mn@doi [\mnras] {10.1093/mnras/stx638}, \href
  {http://adsabs.harvard.edu/abs/2017MNRAS.468.3746C} {468, 3746}

\bibitem[\protect\citeauthoryear{Clark, Plante  \& Greenhill}{Clark
  et~al.}{2012}]{Clark_2012}
Clark M.~A.,  Plante P.~L.,   Greenhill L.~J.,  2012, \mn@doi [International
  Journal of High Performance Computing Applications]
  {10.1177/1094342012444794}, 27, 178

\bibitem[\protect\citeauthoryear{Deller, Tingay, Bailes  \& West}{Deller
  et~al.}{2007}]{Deller_2007}
Deller A.~T.,  Tingay S.~J.,  Bailes M.,   West C.,  2007, \mn@doi
  [Publications of the Astronomical Society of the Pacific] {10.1086/513572},
  119, 318

\bibitem[\protect\citeauthoryear{DuPlain, Ransom, Demorest, Brandt, Ford  \&
  Shelton}{DuPlain et~al.}{2008}]{DuPlain_2008}
DuPlain R.,  Ransom S.,  Demorest P.,  Brandt P.,  Ford J.,   Shelton A.~L.,
  2008, in Bridger A.,  Radziwill N.~M.,  eds, Advanced Software and Control
  for Astronomy {II}. {SPIE}, \mn@doi{10.1117/12.790003}, \url
  {http://dx.doi.org/10.1117/12.790003}

\bibitem[\protect\citeauthoryear{{Escoffier} et~al.,}{{Escoffier}
  et~al.}{2007}]{2007A&A...462..801E}
{Escoffier} R.~P.,  et~al., 2007, \mn@doi [\aap] {10.1051/0004-6361:20054519},
  \href {http://adsabs.harvard.edu/abs/2007A%26A...462..801E} {462, 801}

\bibitem[\protect\citeauthoryear{{Gaensler} \& {Hunstead}}{{Gaensler} \&
  {Hunstead}}{2000}]{2000PASA...17...72G}
{Gaensler} B.~M.,  {Hunstead} R.~W.,  2000, \mn@doi [\pasa] {10.1071/AS00072},
  \href {http://esoads.eso.org/abs/2000PASA...17...72G} {17, 72}

\bibitem[\protect\citeauthoryear{{Green}, {Cram}, {Large}  \& {Ye}}{{Green}
  et~al.}{1999}]{1999ApJS..122..207G}
{Green} A.~J.,  {Cram} L.~E.,  {Large} M.~I.,   {Ye} T.,  1999, \mn@doi [\apjs]
  {10.1086/313208}, \href {http://esoads.eso.org/abs/1999ApJS..122..207G} {122,
  207}

\bibitem[\protect\citeauthoryear{{Green}, {Reeves}  \& {Murphy}}{{Green}
  et~al.}{2014}]{2014PASA...31...42G}
{Green} A.~J.,  {Reeves} S.~N.,   {Murphy} T.,  2014, \mn@doi [\pasa]
  {10.1017/pasa.2014.37}, \href {http://esoads.eso.org/abs/2014PASA...31...42G}
  {31, e042}

\bibitem[\protect\citeauthoryear{{Greenhill}, {Kocz}, {Barsdell}, {Clark}  \&
  {LEDA Collaboration}}{{Greenhill} et~al.}{2014}]{2014era..conf10301G}
{Greenhill} L.~J.,  {Kocz} J.,  {Barsdell} B.~R.,  {Clark} M.~A.,   {LEDA
  Collaboration} 2014, in Exascale Radio Astronomy.

\bibitem[\protect\citeauthoryear{Hunstead \& Gaensler}{Hunstead \&
  Gaensler}{1996}]{Hunstead_1996}
Hunstead R.~W.,  Gaensler B.~M.,  1996, in , Extragalactic Radio Sources.
Springer Netherlands, pp 103--104, \mn@doi{10.1007/978-94-009-0295-4_42}, \url
  {http://dx.doi.org/10.1007/978-94-009-0295-4_42}

\bibitem[\protect\citeauthoryear{{Jankowski} et~al.,}{{Jankowski}
  et~al.}{2015a}]{2015ATel.6903....1J}
{Jankowski} F.,  et~al., 2015a, The Astronomer's Telegram, \href
  {http://adsabs.harvard.edu/abs/2015ATel.6903....1J} {6903}

\bibitem[\protect\citeauthoryear{{Jankowski} et~al.,}{{Jankowski}
  et~al.}{2015b}]{2015ATel.8298....1J}
{Jankowski} F.,  et~al., 2015b, The Astronomer's Telegram, \href
  {http://adsabs.harvard.edu/abs/2015ATel.8298....1J} {8298}

\bibitem[\protect\citeauthoryear{{Jankowski} et~al.,}{{Jankowski}
  et~al.}{2016}]{Jankowski2015c}
{Jankowski} F.,  et~al., 2016, The Astronomer's Telegram, \href
  {http://adsabs.harvard.edu/abs/2015ATel.NNNN....1J} {p.~1}

\bibitem[\protect\citeauthoryear{{Keane} \& {Petroff}}{{Keane} \&
  {Petroff}}{2015}]{KeanePetroff}
{Keane} E.~F.,  {Petroff} E.,  2015, \mn@doi [\mnras] {10.1093/mnras/stu2650},
  \href {http://adsabs.harvard.edu/abs/2015MNRAS.447.2852K} {447, 2852}

\bibitem[\protect\citeauthoryear{Keane, Stappers, Kramer  \& Lyne}{Keane
  et~al.}{2012}]{Keane_2012}
Keane E.~F.,  Stappers B.~W.,  Kramer M.,   Lyne A.~G.,  2012, \mn@doi [Monthly
  Notices of the Royal Astronomical Society: Letters]
  {10.1111/j.1745-3933.2012.01306.x}, 425, L71

\bibitem[\protect\citeauthoryear{{Keane} et~al.,}{{Keane}
  et~al.}{2017}]{Superb1}
{Keane} E.~F.,  et~al., 2017, preprint, \href
  {http://adsabs.harvard.edu/abs/2017arXiv170604459K} {} (\mn@eprint {arXiv}
  {1706.04459})

\bibitem[\protect\citeauthoryear{{Keith} et~al.,}{{Keith}
  et~al.}{2010}]{2010MNRAS.409..619K}
{Keith} M.~J.,  et~al., 2010, \mn@doi [\mnras]
  {10.1111/j.1365-2966.2010.17325.x}, \href
  {http://esoads.eso.org/abs/2010MNRAS.409..619K} {409, 619}

\bibitem[\protect\citeauthoryear{{Kulkarni}, {Ofek}, {Neill}, {Zheng}  \&
  {Juric}}{{Kulkarni} et~al.}{2014}]{2014ApJ...797...70K}
{Kulkarni} S.~R.,  {Ofek} E.~O.,  {Neill} J.~D.,  {Zheng} Z.,   {Juric} M.,
  2014, \mn@doi [\apj] {10.1088/0004-637X/797/1/70}, \href
  {http://esoads.eso.org/abs/2014ApJ...797...70K} {797, 70}

\bibitem[\protect\citeauthoryear{Large, Vaughan  \& Mills}{Large
  et~al.}{1968}]{Large_1968}
Large M.~I.,  Vaughan A.~E.,   Mills B.~Y.,  1968, \mn@doi [Nature]
  {10.1038/220340a0}, 220, 340

\bibitem[\protect\citeauthoryear{Lorimer, Bailes, McLaughlin, Narkevic  \&
  Crawford}{Lorimer et~al.}{2007}]{Lorimer_2007}
Lorimer D.~R.,  Bailes M.,  McLaughlin M.~A.,  Narkevic D.~J.,   Crawford F.,
  2007, \mn@doi [Science] {10.1126/science.1147532}, 318, 777

\bibitem[\protect\citeauthoryear{Lovell}{Lovell}{2008}]{Lovell_2008}
Lovell A.~J.,  2008, in , Science with the Atacama Large Millimeter Array.
Springer Netherlands, pp 191--196, \mn@doi{10.1007/978-1-4020-6935-2_35}, \url
  {http://dx.doi.org/10.1007/978-1-4020-6935-2_35}

\bibitem[\protect\citeauthoryear{{Macquart} et~al.,}{{Macquart}
  et~al.}{2015}]{JPska}
{Macquart} J.~P.,  et~al., 2015, Advancing Astrophysics with the Square
  Kilometre Array (AASKA14), \href
  {http://adsabs.harvard.edu/abs/2015aska.confE..55M} {p.~55}

\bibitem[\protect\citeauthoryear{Manchester, Lyne, Taylor, Durdin, Large  \&
  Little}{Manchester et~al.}{1978}]{Manchester_1978}
Manchester R.~N.,  Lyne A.~G.,  Taylor J.~H.,  Durdin J.~M.,  Large M.~I.,
  Little A.~G.,  1978, \mn@doi [Monthly Notices of the Royal Astronomical
  Society] {10.1093/mnras/185.2.409}, 185, 409

\bibitem[\protect\citeauthoryear{{Masui} et~al.,}{{Masui}
  et~al.}{2015}]{2015Natur.528..523M}
{Masui} K.,  et~al., 2015, \mn@doi [\nat] {10.1038/nature15769}, \href
  {http://esoads.eso.org/abs/2015Natur.528..523M} {528, 523}

\bibitem[\protect\citeauthoryear{Mauch, Murphy, Buttery, Curran, Hunstead,
  Piestrzynski, Robertson  \& Sadler}{Mauch et~al.}{2003}]{Mauch_2003}
Mauch T.,  Murphy T.,  Buttery H.~J.,  Curran J.,  Hunstead R.~W.,
  Piestrzynski B.,  Robertson J.~G.,   Sadler E.~M.,  2003, \mn@doi [Monthly
  Notices of the Royal Astronomical Society]
  {10.1046/j.1365-8711.2003.06605.x}, 342, 1117

\bibitem[\protect\citeauthoryear{{McLaughlin} et~al.,}{{McLaughlin}
  et~al.}{2006}]{2006Natur.439..817M}
{McLaughlin} M.~A.,  et~al., 2006, \mn@doi [\nat] {10.1038/nature04440}, \href
  {http://esoads.eso.org/abs/2006Natur.439..817M} {439, 817}

\bibitem[\protect\citeauthoryear{{Mills} \& {Little}}{{Mills} \&
  {Little}}{1972}]{1972PASAu...2..134M}
{Mills} B.~Y.,  {Little} A.~G.,  1972, Proceedings of the Astronomical Society
  of Australia, \href {http://adsabs.harvard.edu/abs/1972PASAu...2..134M} {2,
  134}

\bibitem[\protect\citeauthoryear{{Murphy}, {Mauch}, {Green}, {Hunstead},
  {Piestrzynska}, {Kels}  \& {Sztajer}}{{Murphy}
  et~al.}{2007}]{2007MNRAS.382..382M}
{Murphy} T.,  {Mauch} T.,  {Green} A.,  {Hunstead} R.~W.,  {Piestrzynska} B.,
  {Kels} A.~P.,   {Sztajer} P.,  2007, \mn@doi [\mnras]
  {10.1111/j.1365-2966.2007.12379.x}, \href
  {http://esoads.eso.org/abs/2007MNRAS.382..382M} {382, 382}

\bibitem[\protect\citeauthoryear{{Nita} \& {Gary}}{{Nita} \&
  {Gary}}{2010a}]{Nita_Gary2010}
{Nita} G.~M.,  {Gary} D.~E.,  2010a, \mn@doi [\pasp] {10.1086/652409}, \href
  {http://adsabs.harvard.edu/abs/2010PASP..122..595N} {122, 595}

\bibitem[\protect\citeauthoryear{{Nita} \& {Gary}}{{Nita} \&
  {Gary}}{2010b}]{2010MNRAS.406L..60N}
{Nita} G.~M.,  {Gary} D.~E.,  2010b, \mn@doi [\mnras]
  {10.1111/j.1745-3933.2010.00882.x}, \href
  {http://esoads.eso.org/abs/2010MNRAS.406L..60N} {406, L60}

\bibitem[\protect\citeauthoryear{{{\"O}zel} \& {Freire}}{{{\"O}zel} \&
  {Freire}}{2016}]{2016ARA&A..54..401O}
{{\"O}zel} F.,  {Freire} P.,  2016, \mn@doi [\araa]
  {10.1146/annurev-astro-081915-023322}, \href
  {http://esoads.eso.org/abs/2016ARA%26A..54..401O} {54, 401}

\bibitem[\protect\citeauthoryear{{Petroff} et~al.,}{{Petroff}
  et~al.}{2015a}]{Petroff_2015}
{Petroff} E.,  et~al., 2015a, \mn@doi [\mnras] {10.1093/mnras/stu2419}, \href
  {http://adsabs.harvard.edu/abs/2015MNRAS.447..246P} {447, 246}

\bibitem[\protect\citeauthoryear{{Petroff} et~al.,}{{Petroff}
  et~al.}{2015b}]{2015arXiv150402165P}
{Petroff} E.,  et~al., 2015b, \mn@doi [\mnras] {10.1093/mnras/stv1242}, \href
  {http://adsabs.harvard.edu/abs/2015MNRAS.451.3933P} {451, 3933}

\bibitem[\protect\citeauthoryear{{Petroff} et~al.,}{{Petroff}
  et~al.}{2016}]{2016arXiv160103547P}
{Petroff} E.,  et~al., 2016, preprint, \href
  {http://esoads.eso.org/abs/2016arXiv160103547P} {} (\mn@eprint {arXiv}
  {1601.03547})

\bibitem[\protect\citeauthoryear{Ransom}{Ransom}{2005}]{Ransom_2005}
Ransom S.~M.,  2005, \mn@doi [Science] {10.1126/science.1108632}, 307, 892

\bibitem[\protect\citeauthoryear{Robertson}{Robertson}{1991}]{Robertson_1991}
Robertson J.,  1991, \mn@doi [Aust. J. Phys.] {10.1071/ph910729}, 44, 729

\bibitem[\protect\citeauthoryear{{Siemion} et~al.,}{{Siemion}
  et~al.}{2012}]{Siemion_2012}
{Siemion} A.~P.~V.,  et~al., 2012, \mn@doi [\apj]
  {10.1088/0004-637X/744/2/109}, \href
  {http://esoads.eso.org/abs/2012ApJ...744..109S} {744, 109}

\bibitem[\protect\citeauthoryear{Smith}{Smith}{2003}]{cite:2}
Smith Q.,  2003, {J}ournal of Homological Model Theory, 7, 1408

\bibitem[\protect\citeauthoryear{{Spitler} et~al.,}{{Spitler}
  et~al.}{2016}]{2016Natur.531..202S}
{Spitler} L.~G.,  et~al., 2016, \mn@doi [\nat] {10.1038/nature17168}, \href
  {http://esoads.eso.org/abs/2016Natur.531..202S} {531, 202}

\bibitem[\protect\citeauthoryear{Staveley-Smith et~al.,}{Staveley-Smith
  et~al.}{1992}]{Staveley_Smith_1992}
Staveley-Smith L.,  et~al., 1992, \mn@doi [Nature] {10.1038/355147a0}, 355, 147

\bibitem[\protect\citeauthoryear{Tate, Garcia  \& Banach}{Tate
  et~al.}{1995}]{cite:0}
Tate Q.,  Garcia L.,   Banach G.,  1995, {A}rchives of the {M}oldovan
  {M}athematical {S}ociety, 0, 78

\bibitem[\protect\citeauthoryear{{Tendulkar} et~al.,}{{Tendulkar}
  et~al.}{2017}]{Tendulkar_2017}
{Tendulkar} S.~P.,  et~al., 2017, \mn@doi [\apjl] {10.3847/2041-8213/834/2/L7},
  \href {http://adsabs.harvard.edu/abs/2017ApJ...834L...7T} {834, L7}

\bibitem[\protect\citeauthoryear{Thornton et~al.,}{Thornton
  et~al.}{2013}]{Thornton_2013}
Thornton D.,  et~al., 2013, \mn@doi [Science] {10.1126/science.1236789}, 341,
  53

\bibitem[\protect\citeauthoryear{Tingay et~al.,}{Tingay
  et~al.}{1995}]{Tingay_1995}
Tingay S.~J.,  et~al., 1995, \mn@doi [Nature] {10.1038/374141a0}, 374, 141

\bibitem[\protect\citeauthoryear{Tingay et~al.,}{Tingay
  et~al.}{2013}]{Tingay_2013}
Tingay S.~J.,  et~al., 2013, \mn@doi [Publ. Astron. Soc. Aust.]
  {10.1017/pasa.2012.007}, 30

\bibitem[\protect\citeauthoryear{Turtle et~al.,}{Turtle
  et~al.}{1987}]{Turtle_1987}
Turtle A.~J.,  et~al., 1987, \mn@doi [Nature] {10.1038/327038a0}, 327, 38

\bibitem[\protect\citeauthoryear{Whiteoak \& Green}{Whiteoak \&
  Green}{1996}]{Whiteoak_1996}
Whiteoak J.~B.,  Green A.~J.,  1996, \mn@doi [Astronomy and Astrophysics
  Supplement Series] {10.1051/aas:1996202}, 118, 329

\bibitem[\protect\citeauthoryear{de Souza, Bunton, Campbell-Wilson, Cappallo
  \& Kincaid}{de~Souza et~al.}{2007}]{de_Souza_2007}
de Souza L.,  Bunton J.~D.,  Campbell-Wilson D.,  Cappallo R.~J.,   Kincaid B.,
   2007, in 2007 International Conference on Field Programmable Logic and
  Applications. pp 62--67, \mn@doi{10.1109/FPL.2007.4380626}

\bibitem[\protect\citeauthoryear{{della Valle}, {Campbell-Wilson}  \&
  {Hunstead}}{{della Valle} et~al.}{1994}]{1994IAUC.6052....2D}
{della Valle} M.,  {Campbell-Wilson} D.,   {Hunstead} R.,  1994, \iaucirc,
  \href {http://esoads.eso.org/abs/1994IAUC.6052....2D} {6052}

\bibitem[\protect\citeauthoryear{van Straten \& Bailes}{van Straten \&
  Bailes}{2011}]{van_Straten_2011}
van Straten W.,  Bailes M.,  2011, \mn@doi [Publ. Astron. Soc. Aust]
  {10.1071/as10021}, 28, 1

\makeatother
\end{thebibliography}

\end{document}